\documentclass[onecolumn, singlespace]{IEEEtran}
\usepackage{cite}
\usepackage{array}
\usepackage{algorithm}
\usepackage{algorithmic}
\usepackage{caption}
\usepackage{longtable}
\usepackage{graphicx,wrapfig}
\usepackage{amssymb}
\usepackage{subfigure}
\usepackage{tikz}
\usepackage{color}
\RequirePackage{afterpage}

\usepackage{amsmath,setspace}
\usepackage{epsfig}

\title{A Cross-layer Contention Based Synchronous MAC Protocol for Transmission Delay Reduction in Multi-Hop WSNs}

\author{Ripudaman~Singh,~\IEEEmembership{}
        Brijesh~K.~Rai,~\IEEEmembership{}
       and~Sanjay~K.~Bose,~\IEEEmembership{Senior Member,~IEEE}
\IEEEcompsocitemizethanks{\IEEEcompsocthanksitem Authors are with the Department
of Electronics and Electrical Engineering, Indian Institute of Technology, Guwahati,
India, 781039.\protect\\
E-mail: {s.ripudaman,  bkrai and skbose}@iitg.ernet.in}
}

\begin{document}

\maketitle
 \thispagestyle{empty}  

\begin{abstract}
  Recently designed cross-layer contention based synchronous MAC protocols like the PRMAC protocol,  for wireless sensor networks (WSNs) enable a node to schedule multi-hop transmission of multiple data packets in a cycle. However, these systems accommodate both the request-to-send data process and the confirmation-to-send data process in the same data transmission scheduling window (i.e. data window). This reduces the length of the multi-hop flow setup in the data window. In a multi-hop scenario, this degrades both the packet delivery ratio (PDR) and the end-to-end transmission delay (E2ETD). In this paper, we propose a cross-layer contention based synchronous MAC protocol, which accommodates the request-to-send data process in the data window and the confirmation-to-send data process in the sleep window for increased efficiency. We evaluate our proposed protocol through ns-2.35 simulations and compare its performance with the PRMAC protocol. Results suggest that in multi-hop scenario, proposed protocol outperforms PRMAC both in terms of the E2ETD and the packet delivery ratio (PDR).
\end{abstract}


\section{INTRODUCTION}
 Prompt detection and reporting of hazards in remote and/or hazardous regions are important applications of wireless sensor networks (WSN). In general, in such applications, event occurrence rate (EOR) is low and non-deterministic. Therefore, most of the time, a sensor node remains in an idle state and most of its limited battery energy is wasted in idle listening. In addition, due to the generally unsafe nature of the monitored region in such applications, a WSN is expected to work after deployment without (or with minimum) human intervention for as long as possible. Therefore, contention based synchronous MAC protocols would be a better choice as the MAC protocol in such applications, since these are typically easier to implement than contention free and hybrid MACs and provide low E2ETD (end-to-end transmission delay) compared to asynchronous MACs.

 In sensor MAC (SMAC) \cite{rd1}, Ye and Heidemann adopt a periodic sleep-wake strategy to reduce the idle listening. For this, SMAC uses a cycle structure which contains three time windows: synchronization window (SW), data transmission scheduling window (DW), and sleep window (SlpW). In SW, each node periodically broadcasts its current sleep-wake schedule to maintain the synchronization. While, in DW, after medium contention with the help of contention window (CW), a node can schedule data packet transmission. In SMAC, after medium contention, a node can schedule only one hop transmission of its data packets, which results in large E2ETD and low packet delivery ratio (PDR) in multi-hop scenarios. In the Routing Enhanced MAC (RMAC) protocol \cite{rd2}, Du et al. utilize cross-layer (routing layer) information, so that, after medium contention, a node can schedule multi-hop transmission of its data packets in DW. For this, RMAC introduces a new control packet, termed as pioneer (PION). PION contains the address of the sender, the next-hop receiver, the previous hop receiver, and the hop distance of the sender from the source of the current flow. For the next and the previous hop receiver, a PION behaves as RTS \cite{rd1} and CTS \cite{rd1}, respectively. With this, in a multi-hop scenario RMAC significantly improve the PDR and E2ETD compared to SMAC. However, in RMAC and other similar works \cite{rd3, rd4,rd5,rd6}, a node can schedule transmission of only one data packet in a cycle, even though the node has multiple data packet to be sent to the same destination. PRMAC \cite{rd7} enables a node to schedule transmission of multiple data packets in a cycle. For this, PRMAC assumes that on the demand of the MAC layer, the routing layer can provide information about the number of data packets in the queue for the requested destination, and can send them irrespective of their order in the queue. With this, PRMAC significantly improves the E2ETD and PDR compared to RMAC. However, PRMAC's performance in a multi-hop scenario is restricted because 1) a node can transmit PION packet, only when the remaining time of DW (or time remains in beginning of SlpW) is more than or equal to the $\text{T}_\text{PION}+\text{SIFS}$, where $\text{T}_\text{PION}$ and $\text{SIFS}$ are the transmission duration of one PION packet and short inter frame space, respectively, and; 2) it accommodates both request-to-send data and confirmation-to-send data process in the DW.

 In this paper, we propose a cross-layer contention based synchronous MAC protocol to reduce the transmission delay for WSNs in a multi-hop scenario. Our proposed protocol uses a novel cycle structure which improves the E2ETD and PDR by increasing the length of the multi-hop flow setup in DW compared to PRMAC, without increasing the duration of DW by putting the request-to-send data process in DW and the confirmation-to-send data process in SlpW.

 The rest of the paper is organized as follows: Section II presents a brief overview of the PRMAC protocol. In Section III, we describe the design details of our proposed protocol, including its cycle structure and data transmission process.  In Section IV, we evaluate our protocol's performance through ns-2.35 simulations based and compare its performance to PRMAC. Finally, we conclude our work in Section V.

\section{Overview of PRMAC}
\begin{figure} [htbp!]
    \centering
    \label{fig1}
    \includegraphics[scale=1.6]{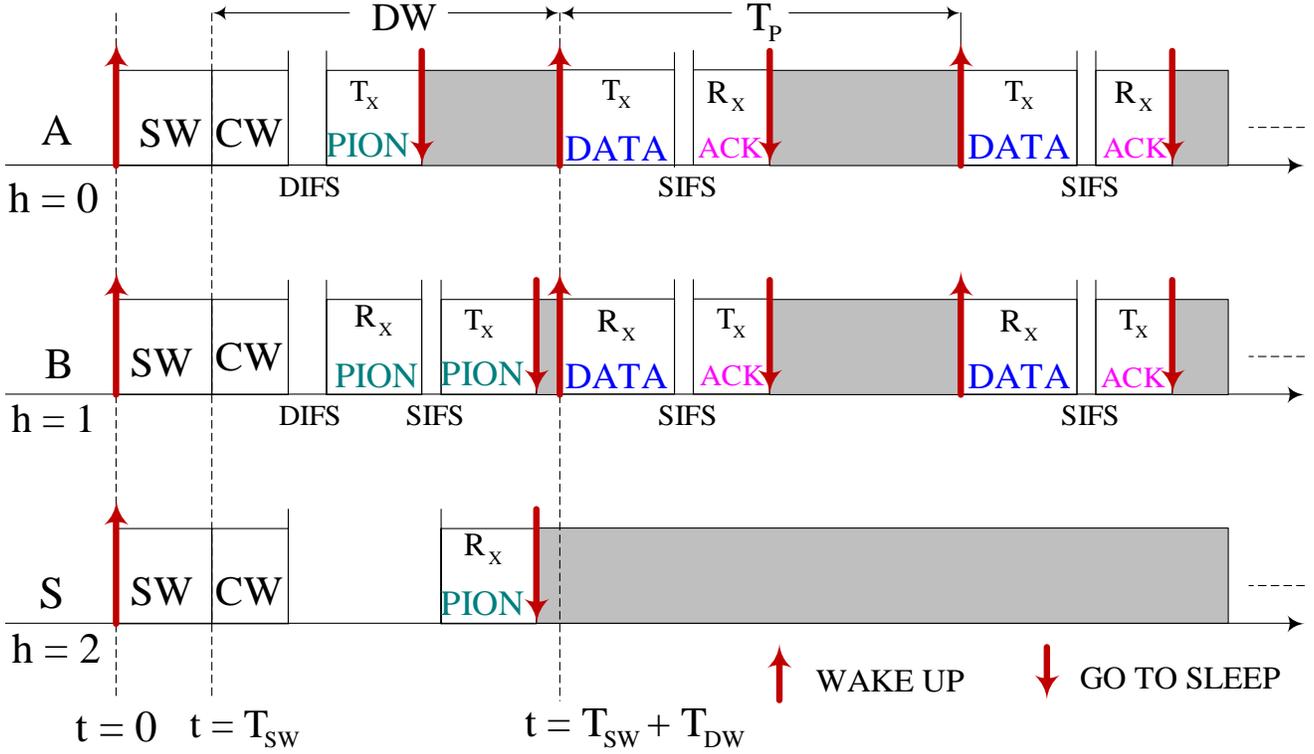}\vspace{-2.0mm}
    \caption{\footnotesize Data Transmission Process of PRMAC.}
    \label{Fig. 1}
\end{figure}

For the sake of completeness, we provide here a brief overview of the data transmission process of PRMAC. As in other existing cross-layer contention based synchronous MAC protocols, PRMAC follows the cycle structure proposed by SMAC. Data transmission process of PRMAC is shown in Fig. 1, assuming that node A has two data packets to send to the two-hop distant sink S, and B is the next-hop receiver of A. In DW, after medium contention, the node A sends a PION packet to forward its data transmission request to its next-hop receiver B. PION transmitted by the node A contains following information: 1) A wants to send two data packets to the node B, and; 2) A is the source of the current flow, i.e., A lies at 0 hop distance from the source of the current flow. The node B transmits a PION to send the data transmission request to the node S, and to inform the node A that it can receive data packets. The PION transmitted by the B contains following information: 1) the number of data packets it can receive from the node A; 2) the number of data packets it wants to send to S, and; 3) hop distance of the node B from the source of current flow (i.e., the node A). After receiving confirmation, the node A goes in sleep. Since the remaining time of DW is not sufficient to transmit a PION packet (i.e., time remains in the beginning of SlpW is less than the $\text{T}_\text{PION}+\text{SIFS}$), B and S also go in sleep.

At the beginning of SlpW, A and B wake up to transmit and receive the first data packet, respectively. Since, the node B doesn't receive confirmation to its request from S, therefore, along with A, it also goes to sleep after transmitting acknowledgement (ACK) packet. After $\text{T}_\text{P}$ (retransmission period) [7] duration from the beginning of SlpW, the nodes A and B wake up to transmit and receive the second data packet, respectively, and then both go in sleep. In this way, both data packets travel one hop towards S in a cycle.

\section{Description of our Proposed Protocol}

%

\subsection{\bf Proposed Cycle Structure}
\begin{figure*} [htbp!]
    \centering
    \label{fig1}
    \includegraphics[scale=1.4]{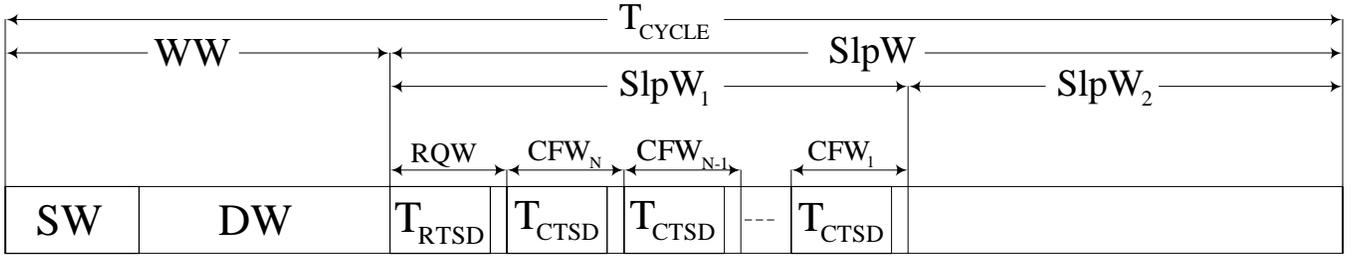}\vspace{-2.0mm}
    \caption{\footnotesize Proposed Cycle Structure.}
    \label{Fig. 1}
\end{figure*}

The cycle structure of our proposed protocol is shown in Fig. 2. Cycle duration is divided into two windows: wake up window (WW) and sleep window (SlpW). In WW, all the network nodes remains in active (wake-up) state. WW is further divided into synchronization window (SW) and data transmission scheduling window (DW). As in \cite{rd1, rd2, rd3, rd4, rd5, rd6, rd7, rd8}, in SW, each node periodically broadcasts a SYNC packet, containing the sender's current sleep-wake schedule, to maintain synchronization. In DW, after medium contention with the help of CW, a node requests its next-hop receiver to receive its data packets and to forward the request further, so that, multi-hop transmission of data packets can be scheduled. For data transmission request, a node transmits RTSD packet \cite{rd8}. In proposed protocol, a RTSD contains the following fields: 1) sender's address; 2) receiver's address; 3) hop distance of the sender from the source of the current flow (i.e., the node who initiates the current flow); 4) number of data packets the sender wants to send to its next-hop receiver, and; 5) the address of final destination. Unlike WW, in SlpW, all the network nodes remain in sleep state. It is further divided into two windows: $\text{SlpW}_{1}$ and $\text{SlpW}_{2}$. $\text{SlpW}_{1}$ contains N $\left( ={\left\lceil {\frac{{{{\rm{T}}_{{\rm{DW}}}}}}{{{\rm{T}}{}_{{\rm{RTSD}}}{\rm{  +  SIFS}}}}} \right\rceil } \right)$ confirmation window (CFW), and one request window (RQW). Here, $\text{T}_\text{RTSD}$ and $\text{T}_\text{DW}$ are the transmission duration of RTSD and duration of DW, respectively, and SIFS is short inter frame space. In Fig. 2, $i^{th}$  $(1\le i \le \text{N})$ CFW is represented by $\text{CFW}_{i}$. Size of RQW and a CFW are equal to the $\text{SIFS} + \text{T}_\text{RTSD}$ and $\text{SIFS} + \text{T}_\text{CTSD}$, respectively, where $\text{T}_\text{CTSD}$ is the transmission duration of a CTSD packet \cite{rd8}. In case, a node receives RTSD containing $i$ as the hop distance of the sender from the source of the scheduled flow, it wakes up at the beginning of $\text{CFW}_{i+1}$ to transmit the CTSD to its upstream node as the response of the received RTSD. The transmitted CTSD contains the sender's address, receiver's address, and number of data packets the node can receive from its upstream node. On the other hand, if a node transmits RTSD with $i$ as the hop distance of the sender from the source of the scheduled flow then it wakes up at the beginning of the $\text{CFW}_{i+1}$ to receive the CTSD from its next-hop receiver. In $\text{SlpW}_{2}$, corresponding to the setup flow, data packets transmission process occurs.

\subsection{\bf Data Transmission Process}
\begin{figure*} [htbp!]
    \centering
    \label{fig1}
    \includegraphics[scale=0.90]{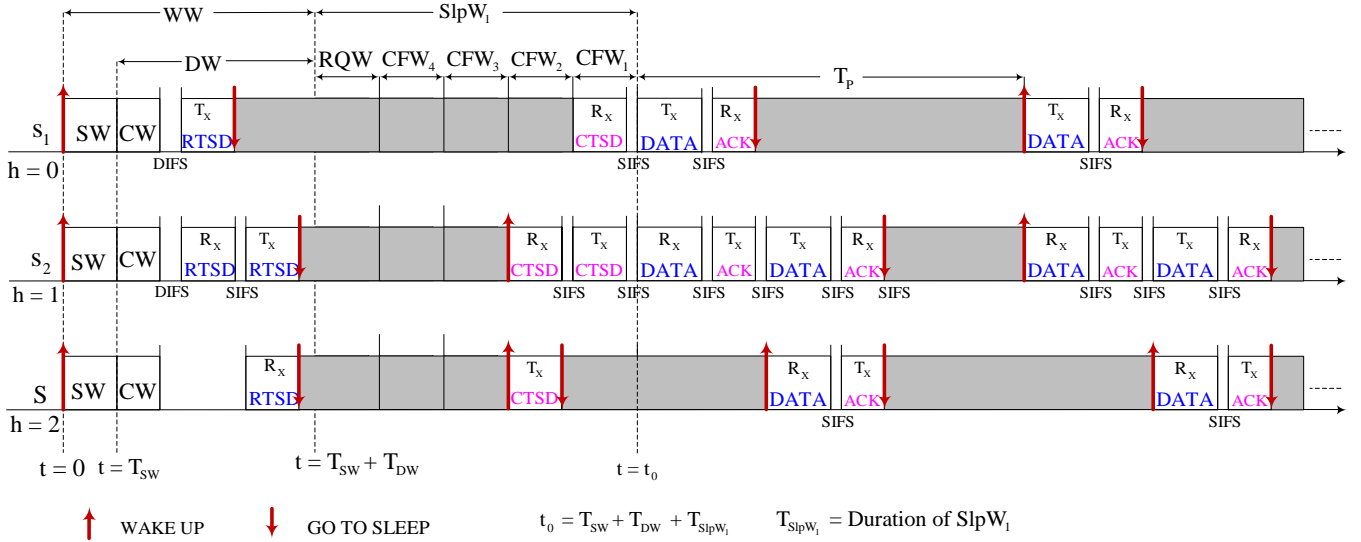}\vspace{-2.0mm}
    \caption{\footnotesize Data transmission process of proposed protocol.}
    \label{Fig. 1}
\end{figure*}
In this subsection, we illustrate the data transmission process of our proposed protocol, assuming that node $s_{1}$ has two data packets to send to the two-hop distant sink S, and $s_{2}$ is the next-hop receiver of $s_{1}$. In its DW, after medium contention, $s_{1}$ sends a data transmission request to its next-hop receiver $\left( s_{2} \right)$ by transmitting a RTSD packet, and then goes to sleep. This RTSD contains the following: 1) address of the node $s_{1}$ as the sender; 2) address of node $s_{2}$ as the receiver; 3) address of node S as the final destination; 4) number of data packets $s_{1}$ wants to send, and; 5) 0 as the hop distance of $s_{1}$ from the source of the current flow. (Note that $s_{1}$ is the source of the current flow in Fig. 3). In case, the node $s_{2}$ can receive data packet from $s_{1}$, $s_{2}$ sends a data transmission request to the node S, which is its next-hop receiver corresponding to the final destination in received RTSD. As done for the node $s_{1}$, after transmitting RTSD $s_{2}$ goes in sleep. RTSD transmitted by $s_{1}$ contains following: 1) address of node $s_{2}$ as the sender; 2) address of node S as the next-hop receiver; 3) address of S as the final destination; 4) number of data packets $s_{2}$ has to send to S, and; 5) 1 as the hop distance of sender from the source of current flow (i.e., $s_{1}$).

After receiving RTSD, S goes in sleep. In case, it can receive data packets from $s_{2}$, it wakes up at the beginning of $\text{CNF}_{2}$, and send a CTSD to the node $s_{2}$. The CTSD sent by S contains the number of data packets S can receive from $s_{2}$. Similarly, at the beginning of $\text{CNF}_{1}$, node $s_{1}$ sends a CTSD to inform the node $s_{1}$ about the number of data packets it can receive. In this way, in proposed protocol, accommodation of request-to-send data (RTSD) and confirmation-to-send data (CTSD) process in two different time windows results in the increase in the scheduled flow length by one hop compared to PRMAC, without increasing $\text{T}_\text{DW}$.

Nodes $s_{1}$ and $s_{2}$ wake up at the beginning of $\text{SlpW}_{2}$ to transmit and receive the first data packet, respectively. After receiving data packet from $s_{1}$, $s_{2}$ forwards the received data packet to the node S. After $\text{T}_\text{p}$ (retransmission period) \cite{rd7} duration from the beginning of SlpW2, $s_{1}$ and $s_{2}$ wake up to transmit and receive second data packet, respectively. Then $s_{2}$ forward the received data packet to the sink S. In this way, in the proposed protocol, data packets travel from source $s_{1}$ to destination S in one cycle. (For the same example, PRMAC would require two cycles as in Fig. 1).

\section{\bf Performance Evaluation}
We evaluate the performance of our proposed protocol using the ns-2.35 simulator. We compare our proposed protocol with PRMAC. For performance evaluation, we have randomly deployed $900$ sensor nodes in a uniform fashion to monitor an area of $1800 \text{m} \times 1800 \text{m}$, and have placed a sink at the center $(900 \text{m}, 900\text{m})$. For performance analysis, we vary the hop distance $\left( h \right)$ between the source and sink. In our simulations, $h$ is varied from 1 to 6 hops. For each value of $h$, we randomly choose a sensor node, among the sensor nodes lying at $h$ hop distance from the sink, as the source node. This chosen source node generates constant bit rate (CBR) traffic with packet arrival interval (PAI) 1.0s. In addition to this, in our simulations, each node is assumed to have one omnidirectional antenna and uses the two ray ground reflection radio propagation model.  As in \cite{rd2, rd9},  we assume that each node is aware of its next-hop receiver as per the shortest path and all the nodes are following the same sleep-wake schedule.  Frame size parameters are given in Table I; cycle duration related parameters of proposed and PRMAC protocol are given in Table II and III, respectively. Networking parameters are similar to that of PRMAC \cite{rd7}.

\begin{table}[h!]
\caption{Frame Sizes}\vspace{-5mm}
\begin{center}
\begin{tabular}{|p{3.0cm}|p{0.4cm}||p{3.0cm}|p{0.4cm}|}
  \hline
  \fontsize {10}{7}\selectfont{Frame Type}  & \fontsize {10}{7}\selectfont{Size}  & \fontsize {10}{7}\selectfont{Frame Type}& \fontsize {10}{7}\selectfont{Size}\\
  \hline
  \fontsize {8}{7}\selectfont{RTSD (Proposed)}  & \fontsize {8}{7}\selectfont{12 bytes}  & \fontsize {8}{7}\selectfont{CTSD (proposed)}& \fontsize {8}{7}\selectfont{12 bytes}\\
  \hline
  \fontsize {8}{7}\selectfont{DATA (PRMAC/Proposed)}  & \fontsize {8}{7}\selectfont{50 bytes}  & \fontsize {8}{3}\selectfont{PION (PRMAC)} &\fontsize {8}{7}\selectfont{14 bytes}\\
  \hline
  \fontsize {8}{7}\selectfont{ACK (Proposed/PRMAC)}  & \fontsize {8}{7}\selectfont{10 bytes}  & \fontsize {8}{7}\selectfont{SYNC (PRMAC/Proposed)} & \fontsize {8}{7}\selectfont{9 bytes}\\
  \hline
\end{tabular}%
\end{center}
\end{table}
%
\begin{table}[h!]
\caption{Cycle Duration Parameters}\vspace{-5mm}
\begin{center}
\begin{tabular}{|p{1.1cm}|p{1.0cm}|p{1.0cm}|p{1.0cm}|p{0.8cm}|p{0.8cm}|}
  \hline
  \fontsize {8}{7}\selectfont{Protocol}  & \fontsize {6}{5}\selectfont{SW (ms)}  & \fontsize {6}{5}\selectfont{DW (ms)}& \fontsize {6}{5}\selectfont{$\text{SlpW}_{1}$ (ms)}& \fontsize {6}{5}\selectfont{$\text{SlpW}_{2}$ (s)} & \fontsize {6}{5}\selectfont{$\rm{T_{CYCLE}}$ (s)}\\
  \hline
  \fontsize {8}{7}\selectfont{Proposed}  & \fontsize {8}{7}\selectfont{55.2 }  & \fontsize {9}{3}\selectfont{117.0} &\fontsize {8}{7}\selectfont{500.0} & \fontsize {8}{7}\selectfont{14.3278} & \fontsize {8}{7}\selectfont{15.0}\\
  \hline
\end{tabular}%
\end{center}
\end{table}
\begin{table}[h!]
\caption{Cycle Duration Parameters}\vspace{-5mm}
\begin{center}
\begin{tabular}{|p{1.1cm}|p{1.2cm}|p{1.2cm}|p{1.4cm}|p{1.2cm}|}
  \hline
  \fontsize {8}{7}\selectfont{Protocol}  & \fontsize {6}{5}\selectfont{SW (ms)}  & \fontsize {6}{5}\selectfont{DW (ms)}& \fontsize {6}{5}\selectfont{$\text{SlpW}$ (ms)} & \fontsize {6}{5}\selectfont{$\rm{T_{CYCLE}}$ (s)}\\
  \hline
  \fontsize {8}{7}\selectfont{PRMAC}  & \fontsize {8}{7}\selectfont{55.2}  & \fontsize {9}{7}\selectfont{117.0}  & \fontsize {8}{7}\selectfont{14.8278} & \fontsize {8}{7}\selectfont{15.0}\\
  \hline
\end{tabular}%
\end{center}
\end{table}

\begin{figure*}[htbp!]
 \centering
   \subfigure[]{
  \includegraphics[scale=0.41]{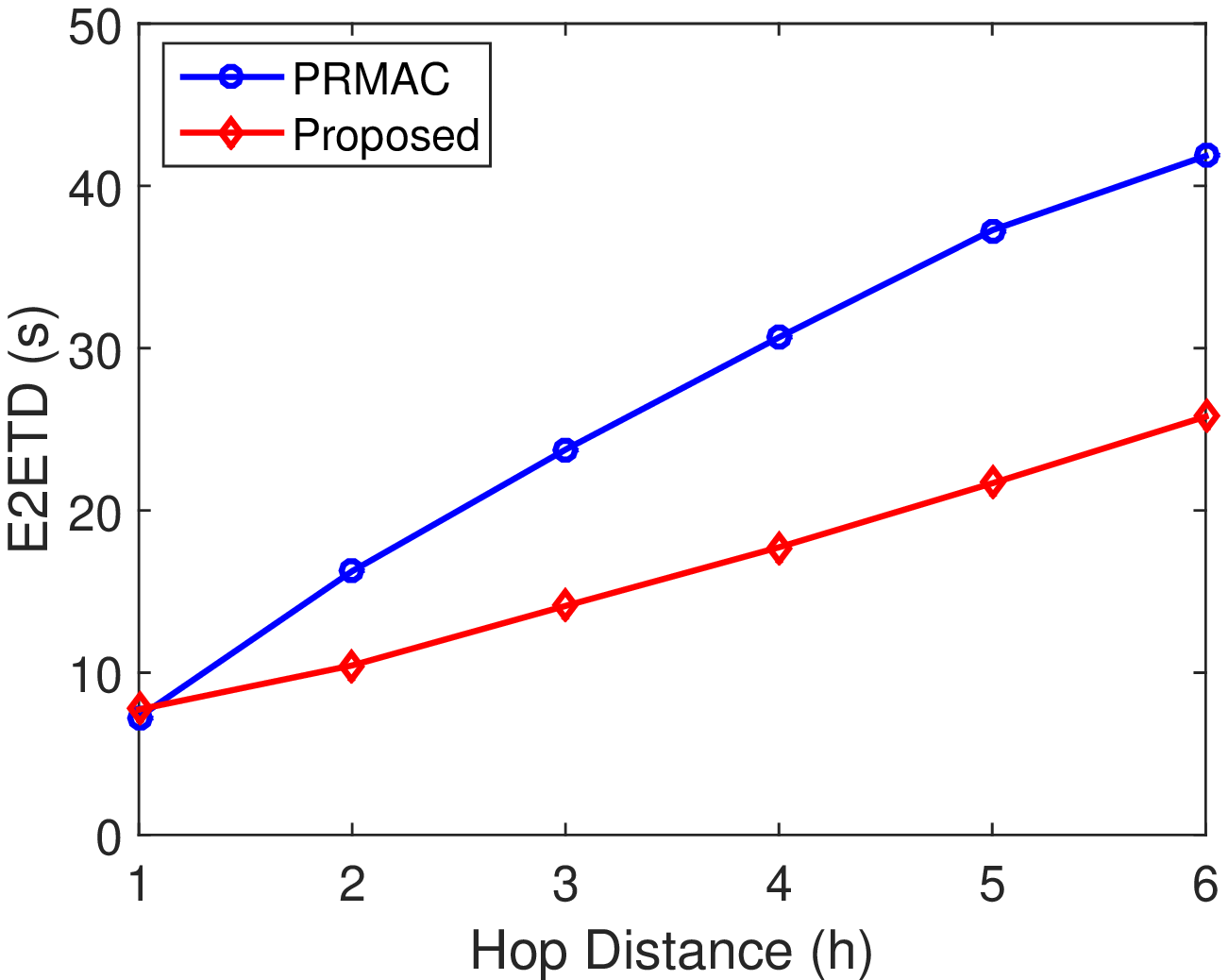}
   \label{fig:subfig2}\vspace{-4mm}
   }\hspace{-2.0em}
   \subfigure[]{
  \includegraphics[scale=0.41]{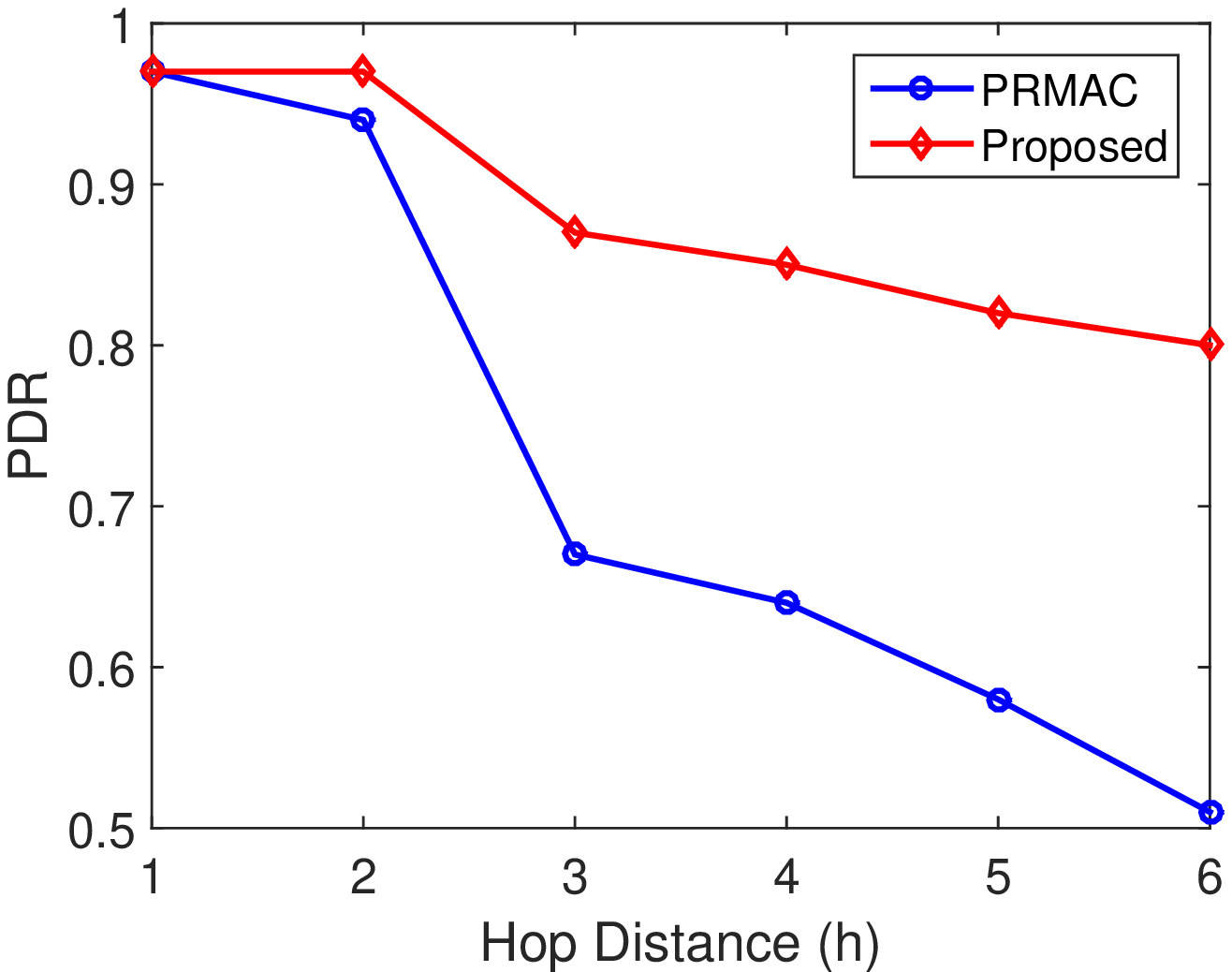}
   \label{fig:subfig2}
   }\hspace{-2.0em}
   \subfigure[]{
  \includegraphics[scale=0.41]{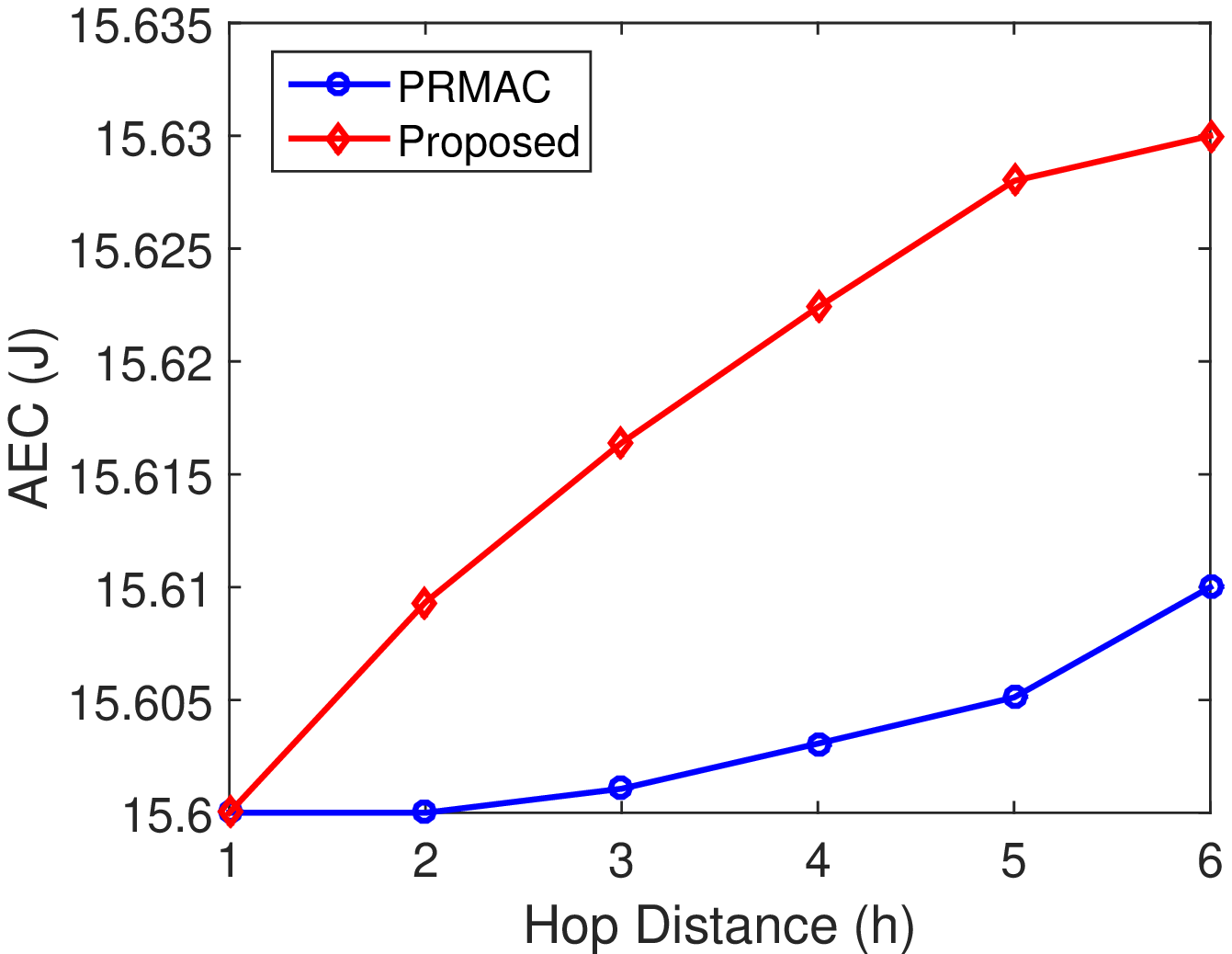}
   \label{fig:subfig2}
   }
 \label{fig:subfigureExample}\vspace{-2.0mm}
 \caption[Optional caption for list of figures]{
  Performance Comparison.}
\end{figure*}

Fig. 4 (a), Fig. 4 (b), and Fig. 4(c) show comparison of E2ETD, PDR,  and average energy consumption (AEC) of PRMAC and proposed protocol. In Fig. 4 (a), Fig. 4 (b), and Fig. 4 (c) results are averaged over $60$ simulations with different seeds, each lasting for $300\text{s}$.
E2ETD is defined as the difference between the reception time of the first data packet at the sink and generation time of the same data packet at the source node. As shown in Fig. 4 (a), with increasing $h$, the difference between the E2ETD of PRMAC and proposed protocol increases. This occurs because, unlike PRMAC, the proposed protocol accommodates the request-to-send and confirmation-to-send data processes in two separate time windows.  Moreover, the proposed protocol allows a node to send a data transmission request even if the remaining time of DW is less than the duration of the RTSD packet.  This can increase the scheduled flow length by up to two hops in the same duration DW in comparison to PRMAC. As a results, the E2ETD of the proposed protocol is less than that of PRMAC. In Fig. 4 (a), in case of $h=6$, E2ETD of proposed protocol is almost $40.0 \%$ less than the PRMAC protocol.

PDR is defined as the ratio of the total number of packets successfully received at the sink to the total packets generated at the source during simulation time. As shown in Fig. 4 (b), the difference between the PDR of PRMAC and that of the proposed protocol increases with increasing $h$. This happens because in the same duration DW, a node can schedule more number of hop for forwarding its data packets in the proposed protocol compared to PRMAC. In Fig. 4 (b), in case of $h=6$, PDR of our proposed protocol is almost $38.0 \%$ more than that of PRMAC.

AEC is defined as the ratio of total energy consumed during the simulation time divided by the number of nodes in the network. From Fig. 4 (c), it is clear that our proposed protocol significantly improves E2ETD and PDR at the cost of very small increase in AEC. In comparison to PRMAC, in proposed protocol, AEC is increased due to the accommodation of request-to-send data process and confirmation-to-send data process in separate time windows.

\section{\bf Conclusions}
In this paper, we proposed a new cross-layer contention based synchronous MAC protocol for WSNs. Our proposed protocol uses a novel cycle structure, which accommodates the request-to-send and confirmation-to-send data processes in two separate time windows, to improve both E2ETD and PDR. Our ns-2.35 simulator based simulation result suggests that the proposed protocol is a better solution than PRMAC as a MAC protocol for WSNs deployed for monitoring the low EOR delay-sensitive events in remote/hazardous regions.

\end{document}